\documentclass[conference]{IEEEtran}
\IEEEoverridecommandlockouts
\usepackage{cite}
\usepackage{amsmath,amssymb,amsfonts}
\usepackage{algorithmic}
\usepackage{algorithm}
\usepackage{multirow}
\usepackage{hyperref}
\usepackage{graphicx}
\usepackage{textcomp}
\usepackage{xcolor}
\def\BibTeX{{\rm B\kern-.05em{\sc i\kern-.025em b}\kern-.08em
    T\kern-.1667em\lower.7ex\hbox{E}\kern-.125emX}}
\begin{document}

\title{SoCRATES: System-on-Chip Resource Adaptive Scheduling using Deep Reinforcement Learning\thanks{
This paper has been accepted for publication by 20th IEEE International Conference on Machine Learning and Applications. The copyright is with the IEEE.}}

\author{\IEEEauthorblockN{Tegg Taekyong Sung}
\IEEEauthorblockA{\textit{EpiSys Science, Inc} \\
Poway, California, United States \\
tegg@episyscience.com}
\and
\IEEEauthorblockN{Bo Ryu}
\IEEEauthorblockA{\textit{EpiSys Science, Inc} \\
Poway, California, United States \\
boryu@episyscience.com}
}

\maketitle

\begin{abstract}
Deep Reinforcement Learning (DRL) is being increasingly applied to the problem of resource allocation for emerging System-on-Chip (SoC) applications, and has shown remarkable promises. In this paper, we introduce SoCRATES (SoC Resource AdapTivE Scheduler), an extremely efficient DRL-based SoC scheduler which maps a wide range of hierarchical jobs to heterogeneous resources within SoC using the Eclectic Interaction Matching (EIM) technique. It is noted that the majority of SoC resource management approaches have been targeting makespan minimization with fixed number of jobs in the system. In contrast, SoCRATES aims at minimizing \textit{average latency} in a steady-state condition while assigning tasks in the ready queue to heterogeneous resources (processing elements). We first show that the latency-minimization-driven SoC applications operate high-frequency job workload and distributed/parallel job execution. We then demonstrate SoCRATES successfully addresses the challenge of concurrent observations caused by the task dependency inherent in the latency minimization objective. Extensive tests show that SoCRATES outperforms other existing neural and non-neural schedulers with as high as 38\% gain in latency reduction under a variety of job types and incoming rates. The resulting model is also compact in size and has very favorable energy consumption behaviors, making it highly practical for deployment in future SoC systems with built-in neural accelerator.
\end{abstract}


\section{Introduction}\label{sec:introduction}
Scheduling is one of the oldest and most universal problems found in our daily lives, and naturally, any tangible gain in efficiency and saving does not go unnoticed. Most systems traditionally have been relying on hand-crafted rules that exhibit limited but acceptable performance in complex and dynamic networks. With a growing number of domains on successfully applying deep learning techniques, it is not surprising to find those neural schedulers to show state-of-the-art performance and dominate hand-crafted algorithms~\cite{mao2018learning,vega2020stomp,avranas2020deep,sung2020deepsocs}.

\begin{figure}[t!] 
\centering
\includegraphics[width=0.44\textwidth]{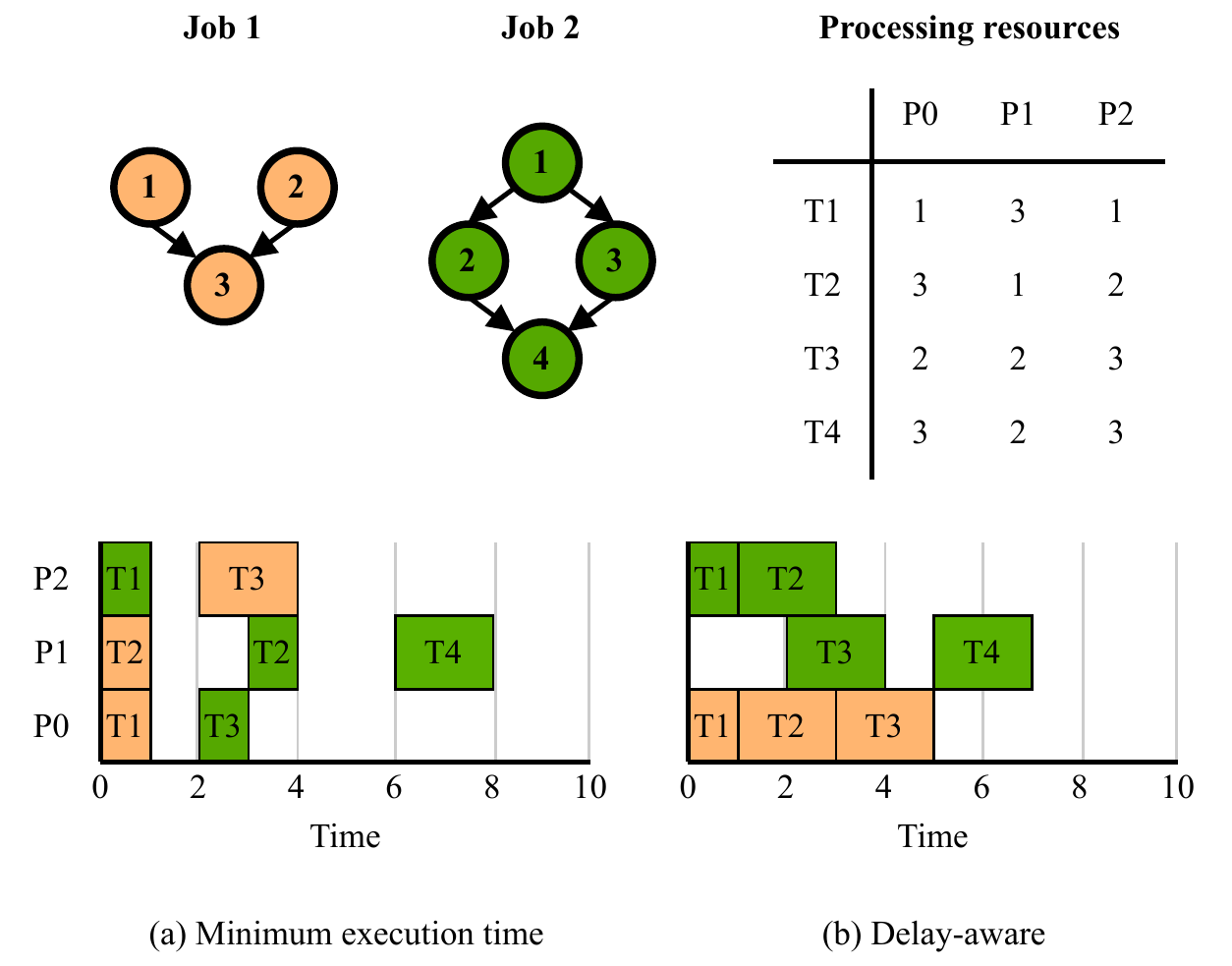}
\vspace{-0.1cm}
\caption{\label{fig:mot_example}  Two different heuristic scheduler examples, each with two disparate incoming job structures: (a) Minimum execution time; (b) Delay-aware.}
\vspace{-0.5cm}
\end{figure}

Despite increasing successes on neural-based schedulers in cluster and network applications, they have limited extensibility due to specific assumptions such as pre-defined job sets with slow injection rates and homogeneous resources. In contrast, this paper investigates scheduling for the System-on-Chip (SoC) application that poses much more challenging aspects such as fast job injection, heterogeneous processing resources, and convoluted (dependent) objectives.

Fig.~\ref{fig:mot_example} depicts two examples of heterogeneous resource scheduling, each with its own optimization objective. While a greedy algorithm picks the minimum executing resources for given tasks (shown in (a)), a delay-aware approach tends to outperform its greedy counterpart by considering data transmission delay (denoted by graph edges). As multiple jobs arrive at the system, increases in job injection rate lead to overlapping tasks, which significantly increases the complexity of action selections. We have found that this elevated complexity makes existing neural SoC schedulers ineffective, as described in Section~\ref{sec:proposed}. Moreover, parallel job execution and hierarchical job graphs cause a scheduling policy asynchrony in action and its outcome. Based on the newfound insights from this analysis, we formulate a novel DRL problem tailored to the SoC application by introducing Eclectic Interaction Matching (EIM) strategy. EIM selects the irregularly injected task assignments and interlaces the actions with subsequent consequences, thus resolving RL's delayed reward problem. Enabled by the EIM technique and formulation, we develop a new SoC neural scheduling algorithm, {\bf S}ystem-{\bf o}n-{\bf C}hip {\bf R}esource {\bf A}adap{\bf T}iv{\bf E} {\bf S}cheduling (SoCRATES). To our knowledge, SoCRATES is the first neural scheduler capable of handling fast job injection rates, diverse job types, and heterogeneous computing resources, all simultaneously with the delay minimization objective.

The performance of SoCRATES is extensively evaluated using a high-fidelity Domain-Specific SoC Simulator (DS3), which injects jobs at high frequencies and maps them to multiple processing resources according to the scheduler. Under a wide range of job injection frequencies and job types, SoCRATES outperforms both non-neural (heuristic) and emerging neural algorithms measured in terms of average latency, job completion time, and energy efficiency. Moreover, the resulting model is compact in size and yields lower energy consumption than other schedulers, even if energy was not part of the optimization objective. 

The major contributions of this work are the following:
\begin{itemize}
    \item Introduction of new DRL design challenges for a high-fidelity SoC chip scheduling simulation environment
    \item Formulation of state, action, and reward statements tailored to a streaming job scheduling simulation
    \item Introduction and application of Eclectic Interaction Matching strategy for addressing variable action sets by redistributing returns according to respective time-varying agent experiences
    \item Feasibility demonstration via increased performance compared to existing heterogeneous resource schedulers
    \item Release of code and models at \url{https://github.com/EpiSci/SoCRATES}.
\end{itemize}

\section{System-on-chip simulation} \label{sec:soc_env}
DS3 is recently proposed for a high-fidelity SoC environment emulated in a heterogeneous SoC computing platform~\cite{arda2020ds3,sung2021gymds3}. DS3 integrates comprehensive system-level design features on heterogeneous resources or processing elements (PEs)\footnote{Resources denote the profiling information, and processing elements indicate the workloads emulated in the simulation kernel.} and provides scheduling and power-thermal management design components. DS3 executes a clock signal that is generated by the clock frequency. The essential characteristic is that DS3 continuously generates indefinite jobs at every stochastic clock signal. In addition, its job generator, distributed PEs, and simulation kernel, all of which are executed in parallel, share the same clock signal. Our goal is to minimize \textit{average latency},
\begin{equation}
    L = \frac{\text{number of completed jobs}}{\text{cumulative execution time}}
\label{eq:latency}
\end{equation}
The cumulative execution time indicates the summation of time from the beginning in simulation to final job completions. Since DS3 is executed in a non-preemptive setting, we do not consider task preemption.

\subsection{Workload characteristic}\label{sec:soc_env:workload}
A job in DS3 may represent a collection of interleaved tasks defined for real-world applications such as wireless communication and radar processing. To resemble them, we define a job structure as a directed acyclic graph (DAG) composed of multiple heterogeneous tasks. Nodes and edges represent task numbers and communication delays, respectively. Job topology expresses task dependency where the scheduler only assigns the tasks free from dependency. Hence, the various task dependency from stochastic job graphs causes variable action sets for a scheduling agent. Heterogeneous resources have different task functionalities (i.e., task execution time, energy/power consumption) and communication bandwidths that compute delays in resource switching. Although DS3 approximates the characteristics of real-world communication or radar processing workloads, we use a synthetic job graph composed of highly varying paths.

\subsection{Simulation characteristics}\label{sec:soc_env:sim}
In general, different types of PEs can be applied based on the resource profile. Each and every PE may have a different task execution time, power consumption, and bandwidth. The scheduling policy's task-PE mapping decisions, along with PE heterogeneity, lead to varying job duration. At every resource switching, delays computed by task transmission cost and PE bandwidth are additionally accrued. In that sense, we find that DS3 exhibits dynamic and realistic operational behaviors. The followings are new DRL design challenges for DS3.
\begin{enumerate}
    \item \textbf{Fast job injection}: DS3 inherently operates with fast job injection rate for having the short critical path and task execution time. The fast job injection rate exponentially increases the feasible number of action sets for various mixes in hierarchical jobs. Further, tasks in different dependencies bring concurrent observations and actions, thus creating delayed reward problems.
    \item \textbf{Heterogeneous resources}: DS3 serves jobs to heterogeneous PEs with different task execution time and communication bandwidth. Edges in the heterogeneous jobs dynamically vary data transmission delays.
    \item \textbf{Hierarchical objectives}: SoC simulation aims not only to minimize average job duration but also to maximize the number of completed jobs~\cite{arda2020ds3}. Unlike makespan minimization for a fixed number of jobs in existing studies, latency minimization inherently resorts to hierarchical optimization. In addition, even if the job DAG topology is defined a priori, the whole input job DAGs in DS3 is still unknown at the beginning of the simulation because indefinite jobs are continuously generated.
\end{enumerate}

\section{Related work} \label{sec:related}
\begin{figure*}[!t] 
\centering
\includegraphics[width=.95\textwidth]{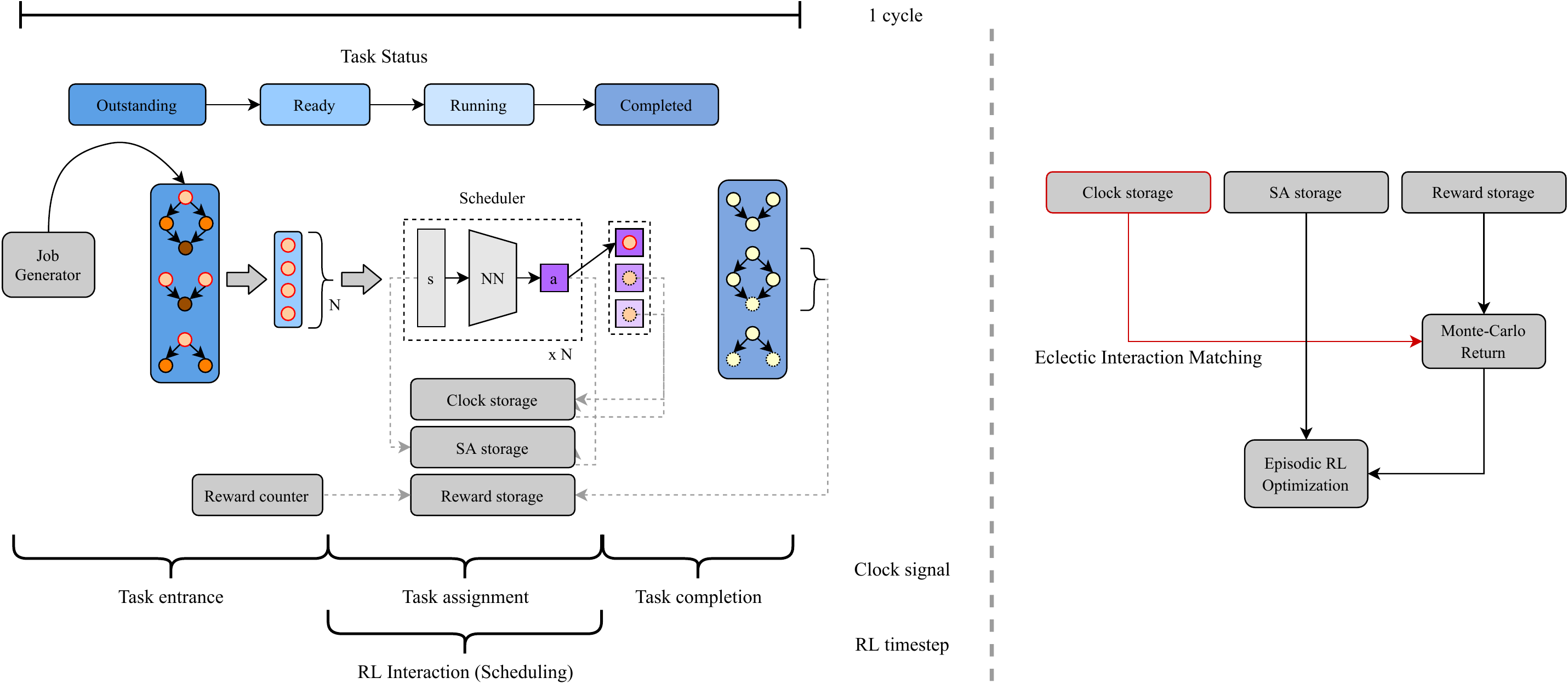}
\caption{\label{fig:overall_struct} An overview of DS3 and SoCRATES. \textbf{Left} Fig. illustrates the SoCRATES scheduler with DS3 workflow. \textbf{Right} Fig. illustrates the training phase, including the Eclectic Interaction Matching.}
\end{figure*}
\noindent \textbf{Rule-based scheduling algorithms.} Traditional schedulers rely almost exclusively on hand-crafted algorithms. First Come First Serve, Earliest Task First (ETF)~\cite{blythe2005task}, Minimum Execution Time (MET)~\cite{buttazzo2011hard}, and Hierarchical Earliest First Time (HEFT)~\cite{topcuoglu1999task} are known to be effective for a wide range of resource management problems. HEFT, for instance, shows remarkable performance in hierarchical job scheduling by first rearranging given tasks based on the importance rank values computed by task computation time and data transmission delays, and greedily mapping ordered tasks to available resources. Also, HEFT uses an insertion-based scheduler that allows filling in the gaps left due to transmission delays associated with previous scheduling decisions. Heuristic scheduling algorithms generally show competitive performance in a simple system, but its performance tends to degrade substantially under scenarios with increased computation and complexity.

\noindent \textbf{DRL-based scheduling algorithms.}
\begin{table}[!t]
\centering
\begin{tabular}{ cccccc } 
\hline
\multirow{2}{*}{Approach} & \multicolumn{3}{c}{Job} & Resource \\
& [CONT] & [HET] & [HIER] & [HET] & \\
\hline
DeepRM~\cite{mao2016resource} & & & & \\
Decima~\cite{mao2018learning} & & \checkmark & \checkmark & \\ 
SCARL~\cite{cheong2019scarl} & & & & \checkmark \\ 
DRM~\cite{sung2019neural} & & \checkmark & \checkmark & \checkmark \\
DeepSoCS~\cite{sung2020deepsocs} & \checkmark & \checkmark & \checkmark & \checkmark \\
SoCRATES (Ours) & \checkmark & \checkmark & \checkmark & \checkmark \\
\hline
\end{tabular}
\caption{A list of DRL-based scheduling methods comparing over the job and resource characteristics. (CONT: Continuous injection, HET: Heterogeneous, HIER: Hierarchical)}
\label{table:sched-comp}
\vspace{-0.8cm}
\end{table}
Table~\ref{table:sched-comp} summarizes a list of emerging DRL-based scheduling algorithms along with their design features. The majority of them corroborates makespan minimization over the fixed number of jobs. DeepRM is the first DRL-based scheduler where the job profile is static and composed of a single-level hierarchy~\cite{mao2016resource}. Its subsequent work, Decima~\cite{mao2018learning}, leverages graph neural networks~\cite{gilmer2017neural} to schedule hierarchical graph-structured jobs to resources adaptively. Decima was developed by placeholder implementation~\cite{abadi2016tensorflow} to tackle stochastic action selection arising from the task dependency. Decima also tested with a heterogeneous resources setting. It, however, mainly provides performance on homogeneous resource scheduling target to cluster application. SCARL~\cite{cheong2019scarl} applies attentive embedding~\cite{vaswani2017attention} to policy networks to maps jobs to heterogeneous resources but simulated in a simple environment~\cite{mao2016resource} and a non-hierarchical workload. Deep Resource Management (DRM)~\cite{sung2019neural} is the first DRL-based scheduler designed for scheduling hierarchical jobs to heterogeneous resources and applied to the SoC chip scheduling simulator. DRM, however, aims to minimize makespan for a single job, which corresponds to a simple scenario and impractical. DeepSoCS~\cite{sung2020deepsocs} extends Decima architecture to schedule SoC jobs in DS3 by applying a heuristic algorithm to map tasks to available resources. DeepSoCS operates with continuously injecting indefinite jobs; however, it shows limited performance under scenarios with fast job injection rates.

\section{Proposed method} \label{sec:proposed}
The systematic workflow of DS3 with a neural scheduling policy is depicted in Fig.~\ref{fig:overall_struct}. The hierarchical jobs are injected into the system with high frequency, where 25 jobs per flop arrive at the job queue. The critical challenges for scheduling in DS3 are that the agent requires to i) immediately perceive continuously generated indefinite job DAGs, ii) adaptively assign a stochastic set of tasks to heterogeneous PEs by considering communication delays and system dynamics, and iii) redistribute returns to actions according to respective time-varying agent experiences. We first provide the RL design pipeline with the state, action, and reward statements tailored to DS3. Afterward, we introduce the EIM strategy for return redistribution. This technique remedies a set of varying actions caused by concurrent observation arising from the task dependency.

\begin{figure*}[!t]
\centering
\includegraphics[width=.95\textwidth]{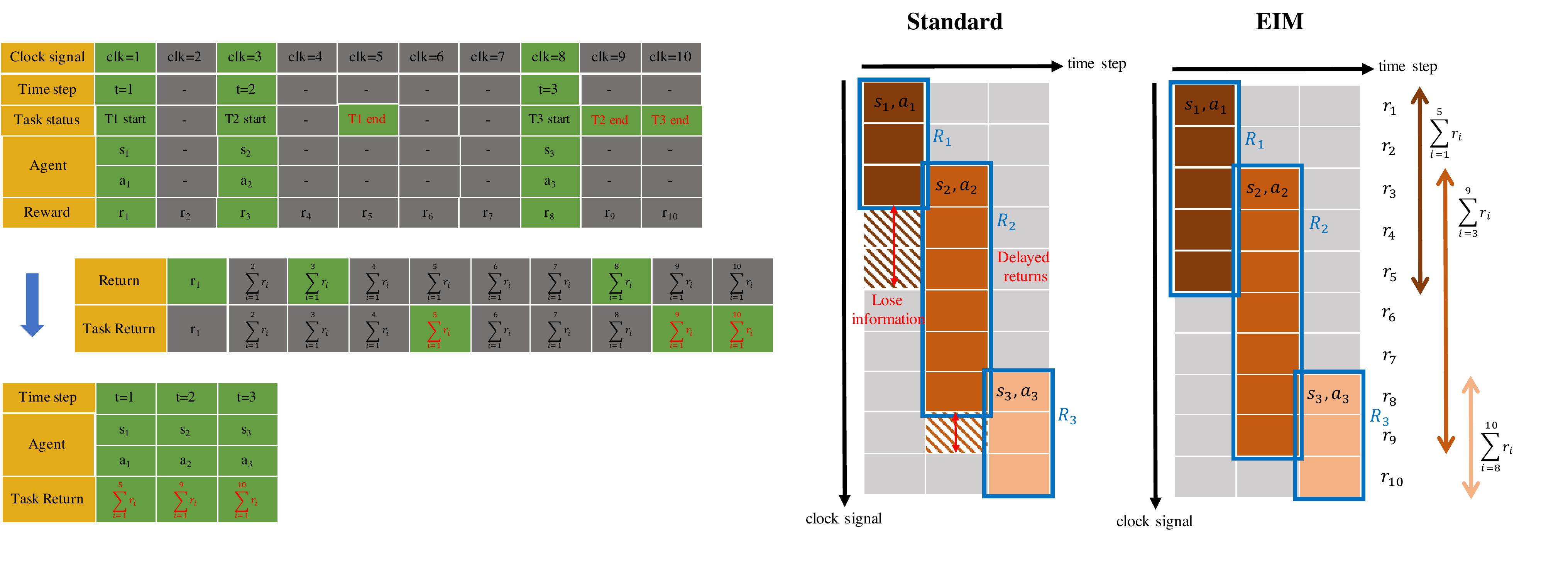}
\vspace{-0.4cm}
\caption{\label{fig:eim} An example of Eclectic Interaction Matching. \textbf{Left} Fig. depicts the process of selecting out the valid information and rearranging the sequential experiments. \textbf{Right} Fig. describes inconsistent interaction and concurrent observation due to the task dependency. Two orthogonal axes show interaction time step and simulation clock signal. EIM strategy selected out assigned task executions accompanied with the returns spanning task duration.}
\end{figure*}
\subsection{Preliminaries} \label{sec:proposed:preliminaries}
RL can be applied in sequential decision-making problems, and its framework assumes Markov decision processes (MDP). The MDP setting can be formalized as a 5-tuple $\langle \mathcal{S}, \mathcal{A}, R, P, \gamma \rangle$~\cite{sutton2018reinforcement}. Here, $\mathcal{S} \in \mathbb{R}^d$ indicates the state space, $\mathcal{A} \in \mathbb{R}^n$ the action space, $R \in \mathbb{R}$ the reward signal that is generally defined over states or state-action pairs. $P: \mathcal{S} \times \mathcal{A} \rightarrow \mathcal{S}$ is a stochastic matrix representing transition probabilities to next states given a state and action. $\gamma \in [0, 1]$ is the discount factor determining how much to care about rewards in maximizing immediate reward myopically or weighing more on future rewards. The objective of RL is to maximize the expected cumulative (discounted) rewards or (discounted) returns. Mathematically, return is calculated by $\mathbb{E}[\sum^T_{t=0} \gamma^{t-1} R(s_{t})]$, where $t$ is the time step for interaction between an agent and an environment. In this paper, we use finite state, finite action, and finite-horizon problem.

\subsection{Agent description} \label{sec:proposed:desc}
\subsubsection{State}
The state representation is designed to capture task/job and resource information. We extract relevant components for multiple overlapping jobs and tasks (i.e., status, remaining dependencies, waiting time). Considering the SoC domain-specific knowledge, we select the dynamic attributes for the observation features. For every task $t$ in job $j$, features are described as follows.
\begin{itemize}
    \item $PE^j_t$ is an assigned PE ID.
    \item $Stat^j_t$ is one-hot embedded task status. Status is classified by one of the labels from ready, running, or outstanding.
    \item $WT^j_t$ is a relative task waiting time from reloaded to the ready status to the current time.
    \item $Pred^j_t$ is the normalized number of remaining predecessors.
    \item $N_{child}$ is the normalized number of all awaiting child tasks in outstanding and ready lists.
    \item $Dep_j$ is the normalized number of hops for the remaining task in task dependency level.
    \item $WT_j$ is a relative job waiting time from injected to the system to the execution time.
\end{itemize}
In a SoC application, each time refers to the simulation clock signal. By concatenating the above features, the environment produces the observation, $\langle (PE^j_t, Stat^j_t, WT^j_t, $ $Pred^j_t)^{T,J}_{t=0,j=0}, (Dep_j, WT_j)^J_{j=0}, N_{child} \rangle$, at each interaction. $J$ denotes the number of job DAGs in job queue, $T$, the number of tasks in job. The job queue holds up to $C$ jobs, where $J \leq C$. The observation serves as base state to the scheduling agent.

\subsubsection{Action}
The number of tasks free from dependency is the agent's action per scheduling decision. The feasible action set is varied due to the hierarchical job graph topology. The agent selects a group of actions $\textbf{a}_t$, where $t$ denotes the interaction time step. Contrarily to explicitly performing variable action sets, this paper iteratively selects an action and individually reassigns outcomes per action assignment. Let the number of tasks in the ready list be $N$. Then, the agent commits an action $a_{t,i} \sim \pi(s_{t,clk})$, $i\leq N$, where $i$ denotes the number of iterations, and $clk$ the temporarily paused clock signal. Finally, we pull out each action value for a PE selection.

\subsubsection{Reward}
The objective in DS3 is to minimize latency which is proportional to the number of completed jobs and inversely to the cumulative execution time as described in ~\eqref{eq:latency}. While online job duration is an adequate reward metric in cluster environments~\cite{mao2018learning}, we empirically discover that negative job duration rewards do not contribute much to latency behavior improvement. Note that cumulative execution time did not significantly differ from simulation length; we neglect the denominator term and approximate the latency to the number of completed jobs. Thereby, based on negative makespan, we incentivize a positive bonus on online job completion.
\begin{equation}
    R(clk) = -0.5\times clk+50\times J_{C}
\label{eq:reward_func}
\end{equation}
Here, $J_C$ denotes the number of newly completed jobs. We carefully design the rewards with a $-0.5$ penalty for every simulation clock signal and $+50$ per job completion.
\begin{figure*}[!t]
\centering
\includegraphics[width=.95\textwidth]{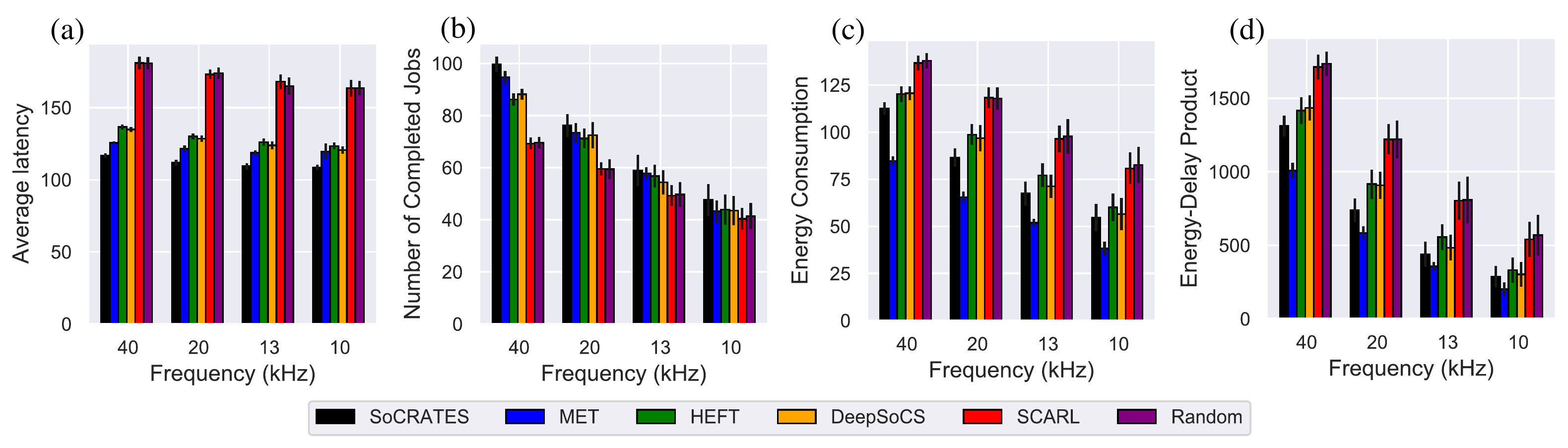}
\caption{\label{fig:overall_res} Performances evaluation on various neural and heuristic schedulers. Experiments are conducted with 40, 20, 13, and 10 kHz frequencies: scales of 25, 50, 75, and 100. For DRL-based schedulers, we evaluate models trained with a scale of 25. DS3 indicates clock signal by flop and energy by joule~\cite{arda2020ds3}. (a) Average latency ($flops/job$). (b) Job completion ($job$). (c) Total energy consumption ($kJ$). (d) Energy-delay product ($kflops \times kJ$).}
\end{figure*}

\subsection{Eclectic Interaction Matching} \label{sec:proposed:eim}
The task-PE mapping in DS3 is done between task entrance execution and task completion; see Fig.~\ref{fig:overall_struct}. In this architecture, the interaction between a scheduling agent and DS3 is fundamentally limited where the agent can access the environment only in the middle of the cycle when the clock is paused. Since the subsequent observation is readily received before the current action has been completed, the task dependencies and different task duration inherently cause concurrent observations. Consequently, the scheduling agent must handle a varying number of stochastic action sets during the action decision stage as jobs are entering the system continuously.

In order to tackle such concurrent observations and a varying number of task mappings, EIM independently counts returns on sequential actions performed at irregular interactions. The left diagram of Fig.~\ref{fig:eim} illustrates this EIM process. First, the agent receives an observation and sequentially selects each action in a varying action set. The clock signal accumulates the negative rewards. Next, the state-action tuples composed with different actions and the clock signal on initial task execution are stored in the buffer. After PEs complete the scheduled tasks, the simulation traverses the clock signal and store the task completion signal in the buffer. The state, action, and task starting clock time are marked with green color on the left diagram, and the task completion clock time is marked with red. Afterward, we compute the returns using reward-by-clock-signal and select out the returns along with the task duration. Thereby, individual tuples in the modified experience sequences shown in the bottom-left diagram of Fig.~\ref{fig:eim} reflect the task duration. Right diagram of Fig.~\ref{fig:eim} denote the return computation using baseline RL return computation~\cite{sutton2018reinforcement} and EIM strategies. The x-axis denotes the standard RL interaction time step, and the y-axis indicates the simulation clock signal. In the standard experience collection strategy, the return is computed at receiving subsequent observation, resulting in partial action consequences are neglected. Based on the completion of the dependent tasks, observations can be concurrently received. In contrast, EIM computes the modified returns for the task duration clock signal, and the calculated returns fully contribute to the action consequence. Hence, each state-action tuple can be paired with the returns that span the task duration.

For training, we use the episodic policy gradient algorithm, actor-critic algorithm~\cite{konda2000actor}, $\theta \leftarrow \theta + \alpha \gamma^t \hat{G}_t^{clk} \nabla_\theta \ln \pi(A_t|S_t;\theta)$. Here, $\theta$ denotes the policy parameter, $\alpha$ the learning rate, and $\gamma$ the discount factor~\cite{sutton1999policy}. $\hat{G}^{clk}_t = (\sum^{CLK}_{clk=0}\gamma^{clk} R_{clk+1})_t$ is an accumulated return computed with EIM strategy. $t$ denotes the task completion clock signal, and $CLK$ the end of simulation length.


\section{Evaluation} \label{sec:eval}

\subsection{Overview} \label{sec:eval:overview}
This section demonstrates the feasibility of SoCRATES using a high-fidelity SoC chip scheduling DS3. We evaluate SoCRATES performance on average latency, job completion, and energy consumption. This paper optimizes the job completion time and uses the DS3 built-in default energy policy to evaluate total energy consumption.

For all experiments, we use Adam optimizer~\cite{kingma2014adam} with learning rate of $3\times10^{-4}$. The gradients are clipped to $1$, and coefficients of value loss and entropy loss are set to $0.5$ and $0.01$, respectively. The simulation continuously generates indefinite jobs at every $clk$ where $clk \sim Exp(\text{scale})$ until the total simulation clock time. A job generator creates five different types of jobs with 20\% chances in each. Each job has a different graph topology and is composed of 10 tasks with different task dependencies. On training, we use a scale of 25 where fast job injection induces the environment more dynamic. An episode runs until 5,000 simulation clock signals, and up to 3 jobs can stack to the job queue. To reduce the training and evaluation time, we conduct all experiments with the pseudo-steady-state condition that the simulation starts with all jobs packed in the job queue~\cite{sung2020deepsocs}. We conduct all experiments with 20 trials using different random seeds.

\subsection{Performance comparison}  \label{sec:eval:performance}
We compare SoCRATES with other schedulers designed for heterogeneous resources in SoC domain and evaluate performances with different job injection rate. We modified existing algorithms in order to run on DS3. For a heuristic scheduler, we modified HEFT such that tasks are dynamically mapped to PE accordingly. Fig.~\ref{fig:overall_res} compares the performance of SoCRATES with that of DeepSoCS~\cite{sung2020deepsocs}, SCARL~\cite{cheong2019scarl}, MET~\cite{braun2001comparison}, and HEFT~\cite{topcuoglu1999task}. Fig.~\ref{fig:overall_res} (a) and (b) compare average latency and number of completed jobs. Latency is the relative metric that depends on the total job completion and total cumulative execution time. A scheduler completing more jobs for a fixed simulation length by fast completion time better performs (the lower, the better). SoCRATES substantially outperforms other schedulers by a large margin in run-time metrics. SoCRATES can generalize to various simulation settings by achieving the lowest latency in all job-injecting frequencies. SCARL~\cite{cheong2019scarl}, initially developed to a single-level job input and a pre-defined number of injecting jobs, cannot effectively schedule continuously injected hierarchical jobs. By observing the indistinguishable performance of SCARL and random scheduling policy, we empirically confirm that effective scheduling of hierarchical task dependency is a critical challenge for SoC scheduling in DS3.

For energy-related performance, Fig.~\ref{fig:overall_res} (c) and (d) show high energy efficiency of the SoCRATES, in spite of not explicitly optimizing for energy-specific metrics. Here, we use the built-in policy to evaluate total energy consumption and energy-delay product (EDP), which is the product of energy consumption and cumulative execution time. Lower energy consumption is preferred for embedded SoC systems, making SoCRATES highly competitive compared to other schedulers.

\section{Conclusion} \label{sec:concl}
This paper presents SoCRATES, a novel DRL-based SoC job/task scheduler with reduced run-time performance and higher energy efficiency over many existing non-neural and neural schedulers. We identify several practical challenges in high-performance SoC scheduler design: (1) complex, inter-mingled objectives; (2) stochastic (non-deterministic) action sets; (3) extremely fast job injection rate; and (4) heterogeneity of incoming jobs and computational resources. We demonstrate that the proposed EIM strategy redistributes returns according to respective time-varying agent experiences, enabling SoCRATES to significantly reduce overall latency, generalize to disparate job injection rates, and adapt to the scaled experiment settings. 

For future work, we plan to conduct a comprehensive set of experiments to further validate the effectiveness and scalability of SoCRATES under a realistic and wide range of jobs. Additionally, we plan to extend SoCRATES to \textit{jointly} optimize for both job completion time and energy efficiency by addressing the increased dimension of the rewards. Finally, we formulate the SoC scheduling as a combinatorial optimization problem, so as to transform it into a mixed-integer programming (MIP) problem. This conversion allows us to apply an emerging neural solver to find optimal task-PE mapping that has the potential to show higher performance than current heuristic MIP solutions.

\bibliographystyle{plain}
\bibliography{main_bib}

\begin{thebibliography}{10}

\bibitem{abadi2016tensorflow}
Mart{\'\i}n Abadi, Paul Barham, Jianmin Chen, Zhifeng Chen, Andy Davis, Jeffrey
  Dean, Matthieu Devin, Sanjay Ghemawat, Geoffrey Irving, Michael Isard, et~al.
\newblock Tensorflow: A system for large-scale machine learning.
\newblock In {\em 12th $\{$USENIX$\}$ symposium on operating systems design and
  implementation ($\{$OSDI$\}$ 16)}, pages 265--283, 2016.

\bibitem{arda2020ds3}
Samet~E Arda, Anish Krishnakumar, A~Alper Goksoy, Nirmal Kumbhare, Joshua Mack,
  Anderson~L Sartor, Ali Akoglu, Radu Marculescu, and Umit~Y Ogras.
\newblock Ds3: A system-level domain-specific system-on-chip simulation
  framework.
\newblock {\em IEEE Transactions on Computers}, 69(8):1248--1262, 2020.

\bibitem{avranas2020deep}
Apostolos Avranas, Marios Kountouris, and Philippe Ciblat.
\newblock Deep reinforcement learning for wireless scheduling with multiclass
  services.
\newblock {\em arXiv preprint arXiv:2011.13634}, 2020.

\bibitem{blythe2005task}
James Blythe, Sonal Jain, Ewa Deelman, Yolanda Gil, Karan Vahi, Anirban Mandal,
  and Ken Kennedy.
\newblock Task scheduling strategies for workflow-based applications in grids.
\newblock In {\em CCGrid 2005. IEEE International Symposium on Cluster
  Computing and the Grid, 2005.}, volume~2, pages 759--767. IEEE, 2005.

\bibitem{braun2001comparison}
Tracy~D Braun, Howard~Jay Siegel, Noah Beck, Ladislau~L B{\"o}l{\"o}ni,
  Muthucumaru Maheswaran, Albert~I Reuther, James~P Robertson, Mitchell~D
  Theys, Bin Yao, Debra Hensgen, et~al.
\newblock A comparison of eleven static heuristics for mapping a class of
  independent tasks onto heterogeneous distributed computing systems.
\newblock {\em Journal of Parallel and Distributed computing}, 61(6):810--837,
  2001.

\bibitem{buttazzo2011hard}
Giorgio~C Buttazzo.
\newblock {\em Hard real-time computing systems: predictable scheduling
  algorithms and applications}, volume~24.
\newblock Springer Science \& Business Media, 2011.

\bibitem{cheong2019scarl}
Mukoe Cheong, Hyunsung Lee, Ikjun Yeom, and Honguk Woo.
\newblock Scarl: Attentive reinforcement learning-based scheduling in a
  multi-resource heterogeneous cluster.
\newblock {\em IEEE Access}, 7:153432--153444, 2019.

\bibitem{gilmer2017neural}
Justin Gilmer, Samuel~S Schoenholz, Patrick~F Riley, Oriol Vinyals, and
  George~E Dahl.
\newblock Neural message passing for quantum chemistry.
\newblock In {\em International Conference on Machine Learning}, pages
  1263--1272. PMLR, 2017.

\bibitem{kingma2014adam}
Diederik~P Kingma and Jimmy Ba.
\newblock Adam: A method for stochastic optimization.
\newblock {\em arXiv preprint arXiv:1412.6980}, 2014.

\bibitem{konda2000actor}
Vijay~R Konda and John~N Tsitsiklis.
\newblock Actor-critic algorithms.
\newblock In {\em Advances in neural information processing systems}, pages
  1008--1014, 2000.

\bibitem{mao2016resource}
Hongzi Mao, Mohammad Alizadeh, Ishai Menache, and Srikanth Kandula.
\newblock Resource management with deep reinforcement learning.
\newblock In {\em Proceedings of the 15th ACM Workshop on Hot Topics in
  Networks}, pages 50--56. ACM, 2016.

\bibitem{mao2018learning}
Hongzi Mao, Malte Schwarzkopf, Shaileshh~Bojja Venkatakrishnan, Zili Meng, and
  Mohammad Alizadeh.
\newblock Learning scheduling algorithms for data processing clusters.
\newblock In {\em Proceedings of the 2019 ACM SIGCOMM Conference}. ACM, 2019.

\bibitem{sung2019neural}
Tegg~Taekyong Sung, Valliappa Chockalingam, Alex Yahja, and Bo~Ryu.
\newblock Neural heterogeneous scheduler.
\newblock {\em arXiv preprint arXiv:1906.03724}, 2019.

\bibitem{sung2020deepsocs}
Tegg~Taekyong Sung, Jeongsoo Ha, Jeewoo Kim, Alex Yahja, Chae-Bong Sohn, and
  Bo~Ryu.
\newblock Deepsocs: A neural scheduler for heterogeneous system-on-chip (soc)
  resource scheduling.
\newblock {\em Electronics}, 9(6):936, 2020.

\bibitem{sung2021gymds3}
Tegg~Taekyong Sung and Bo~Ryu.
\newblock A scalable and reproducible system-on-chip simulation for
  reinforcement learning.
\newblock {\em arXiv preprint arXiv:2104.13187}, 2021.

\bibitem{sutton2018reinforcement}
Richard~S Sutton and Andrew~G Barto.
\newblock {\em Reinforcement learning: An introduction}.
\newblock MIT press, 2018.

\bibitem{sutton1999policy}
Richard~S Sutton, David~A McAllester, Satinder~P Singh, Yishay Mansour, et~al.
\newblock Policy gradient methods for reinforcement learning with function
  approximation.
\newblock Citeseer, 1999.

\bibitem{topcuoglu1999task}
Haluk Topcuoglu, Salim Hariri, and Min-You Wu.
\newblock Task scheduling algorithms for heterogeneous processors.
\newblock In {\em Proceedings. Eighth Heterogeneous Computing Workshop
  (HCW'99)}, pages 3--14. IEEE, 1999.

\bibitem{vaswani2017attention}
Ashish Vaswani, Noam Shazeer, Niki Parmar, Jakob Uszkoreit, Llion Jones,
  Aidan~N Gomez, Lukasz Kaiser, and Illia Polosukhin.
\newblock Attention is all you need.
\newblock {\em arXiv preprint arXiv:1706.03762}, 2017.

\bibitem{vega2020stomp}
Augusto Vega, Aporva Amarnath, John-David Wellman, Hiwot Kassa, Subhankar Pal,
  Hubertus Franke, Alper Buyuktosunoglu, Ronald Dreslinski, and Pradip Bose.
\newblock Stomp: A tool for evaluation of scheduling policies in heterogeneous
  multi-processors.
\newblock {\em arXiv preprint arXiv:2007.14371}, 2020.

\end{thebibliography}

\end{document}